\pdfoutput=1

\documentclass[12pt,a4paper]{article}

\usepackage{jheplike,ifpdf}

\hypersetup{pdftitle={},pdfcreator={},linkcolor=[rgb]{0.15,0.35,0.75},colorlinks=true,citecolor=[rgb]{0.675,0,0.2},urlcolor=[rgb]{0.15,0.35,0.65}}
\newcommand{\eq}[1]{\vspace{-0.5pt}\begin{equation}#1\vspace{-0.5pt}\end{equation}}
\newcommand{\eqs}[1]{\vspace{-0.5pt}\begin{equation}\begin{split}#1\end{split}\vspace{-0.5pt}\end{equation}}

\newcommand{\fwboxL}[2]{\text{\makebox[#1][l]{$#2$}}}
\newcommand{\fwboxR}[2]{\text{\makebox[#1][r]{$#2$}}}
\renewcommand{\hat}{\widehat}

\title{~\\[140pt]{\LARGE \mbox{The Conformal BMS Group}}\\[-20pt]}
\author[\dagger]{\vspace{-20pt}Sasha~J.~Haco}
\author[\dagger]{\!\!\!,\,\,Stephen~W.~Hawking}\author[\dagger]{\!\!\!,\,\,Malcolm~J.~Perry}\affiliation[\dagger]{Department of Applied Mathematics and Theoretical Physics, Centre for Mathematical Sciences, Wilberforce Road, University of Cambridge, Cambridge, CB3 0WA, UK}
\author[\star]{\!\!\!,\,\,Jacob~L.~Bourjaily}\affiliation[\star]{Niels Bohr International Academy and Discovery Center, University of Copenhagen,\\The Niels Bohr Institute, Blegdamsvej 17, DK-2100, Copenhagen \O, Denmark}
\abstract{
We describe the conformal symmetries of asymptotically flat spacetime. These represent an extension of the BMS group that we call the {\it conformal BMS group}. Its general features are discussed. 
}

\begin{document}\maketitle
\vspace{-0pt}\section{Introduction}\label{sec:introduction}\vspace{-8pt}
In four-dimensional Minkowski space, the isometries of the spacetime are given by the ten independent solutions to Killing's equation. These solutions allow one to form the Poincar\'e algebra, made up of four translations in each of the spacetime directions, plus three boosts and three spatial rotations. These are the symmetries of special relativity.

As soon as gravitational fields are included via general relativity, the standard isometry transformations of flat space must be revised. In the 1960s Bondi, van der Burg, Metzner and Sachs (BMS) postulated that there must be some way in which the full Poincar\'e group represent `approximate' symmetry transformations \cite{Sachs:1962zza}, \cite{Bondi:1962px}. They studied these approximate symmetries of curved spacetime by investigating the asymptotic symmetries of asymptotically flat spacetimes at null infinity---if the spacetime were asymptotically flat, then infinitely far away from any gravitational fields we must in some sense be able to reproduce the Poincar\'e group as the symmetry group. This group of asymptotic symmetries is known as the BMS group \cite{Sachs:1962zza}, \cite{Bondi:1962px}, a larger group than the Poincar\'e group of flat space, that consists of the ordinary Lorentz transformations plus an infinite number of `supertranslations'.

This BMS group has been extensively studied over the years. Penrose investigated the BMS group as a symmetry group on null infinity \cite{Penrose:1962ij}, and later with Newman, he looked into possible subgroups of BMS that might arise when considering scattering problems and the emission of radiation out to infinity \cite{Newman:1966ub}.

More recently, the BMS group has received renewed attention. An extension to the BMS group has been proposed to include `superrotations' \cite{Barnich:2011ct,Barnich:2010eb,Barnich:2009se}, and work has been done on the conserved quantities that would be associated to the asymptotic symmetries of the BMS group \cite{Hawking:2016b,Wald:1999wa,Barnich:2011mi,Flanagan:2015pxa,Compere:2016hzt}. In the quantum picture, these conservation laws amount to relations between ingoing and outgoing scattering states \cite{Kapec:2014opa,Strominger:2013jfa}, and have been shown to be equivalent to so-called soft theorems \cite{He:2014laa,Strominger:2014pwa}, and subleading soft-theorems \cite{Lysov:2014csa}, originally formulated by Weinberg and Low \cite{Low:1954kd,Weinberg:1965nx}. Within the last year, the effect of these symmetries on black hole spacetimes has been investigated, and the potential for these conservation laws to provide answers to the black hole information paradox \cite{Hawking:1976ra,Hawking:2016msc}.

While the Poincar\'e and BMS groups describe the symmetries of special and general relativity, for any theory that also admits a conformal symmetry, the necessary group of isometries must be larger. In flat space, the Poincar\'e group gets extended to the conformal group at spacelike infinity, and at null infinity, one needs not the BMS group but a conformal version of it, which is developed here. 

Conformal symmetry is at the heart of many important physical theories. For example, Maxwell's free field equations are conformally invariant, as is the massless Dirac equation. In terms of gravity, the situation is less clear, but for empty space, the Weyl tensor is unchanged by conformal transformations to the metric \cite{Penrose:1965am}.  Another hint at conformal symmetry in gravity is through the connection with Yang-Mills theory: some aspects of gravity, particularly scattering amplitudes, can be regarded as the product of two Yang-Mills theories \cite{Borsten:2015pla} - and we know Yang Mills to be a classically conformally invariant theory in Minkowski space.

Given that $\mathcal{N}\!=\!4$ Yang-Mills theory exhibits conformal symmetry, an obvious next step will be to study the action of the {\it conformal BMS group} in this context. The BMS group has previously been shown to be a conformal extension of the Carroll group \cite{Duval:2014-2,Duval:2014lpa}. A generalization of the BMS group for supergravity has also been studied \cite{Awada:1985by}, although without investigation into asymptotically conformal transformations. Recently, work on classifying the asymptotic symmetry algebras of theories in different dimensions has been studied in the context of holography \cite{Irakleidou:2016xot}. 

This paper is organized as follows. In \mbox{section \ref{sec:flat_space_symmetries}} we review the well known symmetry groups of Minkowski space, and extend this to the asymptotic symmetries of the BMS group in \mbox{section \ref{sec:asymptotically_flat_space_symmetries}}. In \mbox{section \ref{sec:conformal_bms}} we introduce the conformal BMS group, and discuss its algebra and properties. The closure of this algebra is more subtle than it may at first appear---due to the fact that the generators are metric-dependent. This is achieved through a modified bracket, defined and discussed in \mbox{section \ref{subsec:modified_bracket}}. We illustrate this modified bracket algebra with a detailed example in \mbox{Appendix \ref{appendix:modified_bracket}}.

\vspace{-6pt}\section{Conformal Symmetries of Flat Space: \\Poincar\'e and Conformal Groups}\label{sec:flat_space_symmetries}\vspace{-8pt}
In four-dimensional flat Minkowski spacetime, it is possible to identify certain symmetries of the metric---transformations that leave the spacetime invariant. These are the ten isometries which form the well known Poincar\'e group of the symmetries of special relativity. These symmetries are found by asking for which vector fields $\xi$ does the Lie derivative of the metric vanish, in other words, solutions to Killing's equation,
\eq{(\mathcal{L}_\xi g)_{a b}=\nabla_{a} \xi_{b} + \nabla_{b} \xi_{a} = 0.}
$\mathcal{L}_\xi$ is the Lie derivative with respect to the vector field $\xi$. In (3+1)-dimensional Minkowski space, we get ten independent solutions (Killing vectors (KV)) that make up the Poincar\'e group. This Poincar\'e group consists of the Lorentz group, a subgroup made up of three boosts and three spatial rotations, as well as an abelian normal subgroup of four translations in each of the spacetime directions. 

The generators of these symmetry transformations may be written,
\eq{M_{a b}\equiv (x_a\partial_b - x_b\partial_a),\qquad P_a\equiv \partial_a\,,}
where the $M_{a b}$ give the Lorentz transformations and $P_b$ the translations.
The commutation relations are,
\eqs{\left[P_{a}, P_{b} \right] &= 0,
\\ \left[M_{a b}, P_{c} \right] &=  \eta_{b c}\,P_{a} - \eta_{a c}\,P_{b},
\\ \left[M_{a b}, M_{c d} \right] &=  \eta_{a d}\,M_{b c} + \eta_{b c}\,M_{a d} - \eta_{b d}\,M_{a c} - \eta_{a c}\,M_{b d},}
where $\eta_{a b}$ is the Minkowski metric of signature $(-,+,+,+)$. These are the generators of the group $O(3,1)$.

We may also look at transformations which preserve the metric up to a conformal factor,
\eq{\mathcal{L}_\xi \,g = \Omega^2\,g\,.}
By taking the trace, we can solve for $\Omega^2$ and find that the transformations $\xi$ correspond to solutions to the {\it conformal Killing equation}, which in four dimensions is:
\eq{\nabla_{a}\,\xi_{b} + \nabla_{b}\,\xi_{a} - \frac{1}{2}\,g_{a b}\,\nabla_{c}\,\xi^{c} = 0\,.}
The solutions are {\it conformal Killing vectors} (CKV).

In flat space, the conformal Killing vectors consist of the Poincar\'e group, along with an extension to include special conformal transformations generated by $K_\mu$ and dilatations (scalings) generated by $D$:
\eqs{D&\equiv x^a\,\partial_a\,,
\\ K_a&\equiv x^2\,\partial_a - 2\,x_a x^b\,\partial_b\,.\label{d_and_k_definitions}}
The commutation relations are given by:
\eqs{
\left[D , K_{a} \right] &= K_{a},
\\ \left[D , P_{a} \right] &= -P_{a},
\\ \left[K_{a}, P_{b} \right] &= 2 \eta_{a b}D + 2 M_{a b},
\\ \left[K_{a}, M_{b c} \right] &=  \eta_{a b}K_{c} -  \eta_{a c} K_{b}\,.}
These are the generators of the group $O(4,2)$.

\vspace{-6pt}\section{Conformal Symmetries of Asymptotically Flat Space}\label{sec:asymptotically_flat_space_symmetries}\vspace{-8pt}

In a curved spacetime the above transformations no longer hold as exact symmetries. However, in any asymptotically flat spacetime one can define `asymptotic symmetries' which correspond to those transformations that are consistent with the boundary conditions of asymptotic flatness. This amounts to the consideration of an `asymptotic Killing equation'---the solutions to which are known to form a larger group of symmetries, known as the BMS group \cite{Bondi:1962px}. This consists of the ordinary Lorentz transformations, plus an infinite number of `supertranslations' and `superrotations'. Let us briefly review how these symmetries arise in some detail, before extending this algebra to include also the asymptotic manifestations of conformal symmetry. 

Using retarded Bondi coordinates $(u, r, x^A)$, the flat space Minkowski metric is given by,
\eq{ds^2 = - du^2 - 2 du\,dr + r^2 \,\gamma_{AB}\,dx^A dx^B\,,}
where $\gamma_{A B}$ is the unit metric on the two-sphere at infinity. In the Bondi gauge, 
\eq{g_{rr} = g_{rA} = 0,\qquad\,\partial_r \left(\frac{\det(g_{AB})}{r^2} \right) = 0\,.}
In order to maintain this metric asymptotically, any allowed transformations are constrained by a set of boundary conditions. These ensure that any non-zero components of the resulting Riemann tensor have suitable $r$-dependence as $r \to \infty$, so that the curvature falls off sufficiently fast. The corresponding changes to the metric must therefore obey certain fall-off conditions, given by
\eqs{\delta g_{uA} &\sim \mathcal{O}(r^0)\,,
\\ \delta g_{ur} &\sim \mathcal{O}(r^{-2})\,,
\\ \delta g_{uu} &\sim \mathcal{O}(r^{-1})\,,
\\ \delta g_{AB} &\sim \mathcal{O}(r)\,.}
In order to satisfy the Bondi gauge, we also require,
\eq{\delta g_{rr} = \delta g_{rA} = 0,\qquad\,\partial_r \left(\frac{\det(g_{AB} + \delta g_{AB})}{r^2} \right) = 0\,.}
If peeling holds \cite{Newman:1961qr}, any asymptotically flat metric can be written as an expansion in powers of $1/r$. In Bondi coordinates near null infinity, this is,
\eqs{ds^2=&- du^2 - 2 du\, dr + r^2 \,\gamma_{A B}\, dx^A dx^B \\
&+2\frac{m_b}{r} du^2 + r\,C_{AB} dx^A dx^B  + D_{A}\,C^A_{B}\,du\,dx^B +\ldots\,,}
where $D_A$ is the covariant derivative with respect to the metric on the two-sphere, $m_b$ and $C_{AB}$ denote first order corrections to flat space. $m_b$ is the `Bondi mass aspect', and $\partial_u C_{AB} = N_{AB}$ where $N_{AB}$ is the `Bondi news'. Capital letters $A, B,...$ can be raised and lowered with respect to $\gamma_{A B}$.

Transformations that preserve these conditions and therefore maintain the structure of the metric correspond to asymptotic solutions to the Killing equation. These are generated by the vector fields,
\eqs{\xi_T &\equiv f\,\partial_u + \frac{1}{2}D^2 f\,\partial_r - \frac{1}{r} D^A f\,\partial_{A}\,,
\\ \xi_R &\equiv \frac{1}{2}u\, \psi\,\,\partial_u - (\frac{1}{2}r\,\psi - \frac{1}{4}u\,D^2 \psi)\,\partial_r + (Y^A - \frac{u}{2r}D^A \psi)\partial_A\,,}
where $f$ is any scalar spherical harmonic, $Y^A$ are conformal Killing vectors on the 2-sphere, and $\psi\!\equiv\!D_A Y^A$. Further terms that are subleading in $r$ have been neglected. The vectors $\xi_T$ generate infinitesimal `supertranslations' and the $\xi_R$ give the `superrotations'. The supertranslations act to shift individual light rays of null infinity forwards or backwards in retarded time. The standard BMS group of infinitesimal transformations preserving the asymptotically flat metric contains only the the superrotations that are globally well defined on the sphere. These correspond to supertranslations $\xi_T$, and superrotations $\xi_R$ for which $Y^z \!=\! 1, z, z^2$ and its conjugates, when expressed in stereographic coordinates on the two-sphere \cite{Sachs:1962zza}. More recently, an `extended BMS' group has been proposed to include all vector fields $\xi_R$ with $Y^z\!=\! z^{n+1}$ (and conjugates) for any $n$  \cite{Barnich:2010eb,Barnich:2009se,Barnich:2011ct,Compere:2016jwb,Barnich:2011mi}. There is a similar construction at past null infinity.

\vspace{-6pt}\section{The Conformal BMS Symmetry Groups}\label{sec:conformal_bms}\vspace{-8pt}
For the conformal case, we look for asymptotic solutions to the conformal Killing equation, and ask that the infinitesimal changes in the metric satisfy the same fall-off conditions as above.

The group of solutions involves the ordinary BMS supertranslations ($T$) and superrotations ($R$), plus a dilatation ($D$), another sort of conformal dilatation, a \lq BMS dilatation' ($E$), and a new BMS special conformal transformation, a \lq BMS special conformal transformation' ($C$). In our coordinates, at leading order, these are given by,
\eqs{T &\equiv f\,\partial_u + \frac{1}{2}D^2f\,\partial_r - \frac{1}{r}D^A f\,\partial_A\,,
\\ R &\equiv \frac{1}{2}u \psi\,\partial_u  - (\frac{1}{2}r\,\psi - \frac{1}{4}u\,D^2 \psi)\,\partial_r + (Y^A - \frac{u}{2r}D^A \psi)\,\partial_A\,,
\\ D &\equiv u\,\partial_u + r\,\partial_r\,,
\\ E &\equiv \frac{u^2}{2}\,\partial_u + r(u+r)\,\partial_r\,,
\\ C &\equiv \frac{u^2}{4} \zeta\,\partial_u - \left( \frac{u^2}{4} + \frac{r^2}{2} + \frac{u\,r}{2} \right)\zeta\,\partial_r - \frac{u}{2}\left(1 + \frac{u}{2r} \right)D^A\zeta \,\partial_A\,,}
where $\psi\!\equiv\!D_A Y^A$, $\zeta\!\equiv\!D_A Z^A$, and $Y^A$ and $Z^A$ are conformal Killing vectors on the 2-sphere. Note that while the superrotations may be formed from any conformal Killing vectors, the special conformal transformations however vanish if $Z^A$ is a Killing vector. Therefore $C$ is only formed from the divergence of \lq strictly conformal Killing vectors'.

Thus the conformal BMS group is larger than both the conformal group and the BMS group. As well as the infinite number of supertranslations and superrotations, the new special conformal transformation also give an infinite number of symmetries---generated by the infinity of strictly conformal Killing vectors $Z^A$. Just as for the superrotations we can define both global and local special conformal transformations. The conformal BMS group described above is the group $CBMS^+$, as it is defined on future null infinity, $\mathcal{J}^+$. Performing a similar calculation on past null infinity, $\mathcal{J}^-$, we can obtain the corresponding (although different) group $CBMS^-$.

It is also worthwhile considering how the original ({\it i.e.}\ flat space) conformal group fits into this larger asymptotic group. In flat space, there are four special conformal transformations, given by \mbox{equation (\ref{d_and_k_definitions})}. When written in $(u, r, x^A)$ coordinates, these are,
\eqs{K_u &\equiv u^2\,\partial_u + 2 r (u+r)\partial_r\,,
\\ K_r &\equiv 2 u^2\,\partial_u - u^2\partial_r\,,
\\ K_A &\equiv - u (u+2r)\partial_A\,;}
we can thus identify,
\eq{K_u = 2 E\,,}
and the other components are contained within the superrotation and the new special conformal transformation $C$, for suitable choice of $\psi$ and $\zeta$.
For example, choosing coordinates $(u, r, \theta, \phi)$, then
\eqs{Z^{\theta} = - 4 \cot\!\theta,\quad  Z^{\phi} = 0, &\implies \zeta = 4  \implies C + 2 E = K_r\,.}
%

\vspace{-2pt}\subsection{The Modified Bracket}\label{subsec:modified_bracket}\vspace{-2pt}

In order to compute the algebra, there is an important subtlety that must be taken into account: it is not the Lie bracket that is required, but a modified version of it (see {\it e.g.} \cite{Barnich:2011mi}). This is because the vector fields generate perturbations in the metric and these vector fields are themselves metric-dependent. Thus, in calculating the commutator an extra piece must be added or subtracted from the usual bracket in order to take into account how each vector field varies as the metric changes.

Consider the action of a vector field, $\xi_1$ on the metric, followed by another vector, $\xi_2$. We allow metric variations \mbox{$g_{ab}\!\to\!g_{ab}\!+\! \hat{h}_{ab}$} which satisfy the fall-off conditions given above and calculate the possible vector fields, $\xi$, which can give rise to such variations. Thus these vector fields are defined through,
\eq{\hat{\mathcal{L}}_{\xi_1}g = \hat{h}\,,}
where the `conformal' Lie derivative is defined by,
\eq{(\hat{\mathcal{L}}_{\xi}g)_{ab} = \nabla_a \xi_b + \nabla_b \xi_a - \frac{1}{2}g_{ab} \nabla_c \xi^c\,,}
When the vector $\xi_2$ acts on the metric we allow for additional perturbations:
\eqs{\xi_2 &\to \xi_2 + \mu_2\,, 
\\ g + \hat{h} &\to g + \hat{h} + \hat{K}\,}
where $\mu_2$ is a first order perturbation to the vector field and $\hat{K}$ is a second order variation of the metric. We then find the action of $\hat{\mathcal{L}}_{\xi_2}g$ to second order. Explicitly,
\eqs{\hat{K}_{ab} =&\phantom{\,+\,} \mu_2^c\,\partial_c g_{ab} + \xi^c\,\partial_c \hat{h}_{ab} +\,\partial_a \xi^c \hat{h}_{bc} +\,\partial_a \mu_2^c g_{bc} +\,\partial_b \mu^c g_{ac} +\,\partial_b \xi^c \hat{h}_{ac} \\& - \frac{1}{2}g_{ab}\,\partial_c \mu_2^c - \frac{1}{2}\hat{h}_{ab}\,\partial_c \xi^c - \frac{1}{2}g_{ab} \Gamma^c_{\phantom{c}cd} \mu^d - \frac{1}{2}\hat{h}_{ab} \Gamma^c_{\phantom{c}cd} \xi^d - \frac{1}{2}g_{ab} \delta \Gamma^c_{\phantom{c}cd} \xi^d\,,}
where $\delta \Gamma^c_{\phantom{c}cd}$ is the perturbation of the connection $ \Gamma^c_{\phantom{c}cd}$ due to the change \mbox{$g\!\to\!g\!+\!\hat{h}$}. Asking that the corresponding changes to the metric still satisfy the boundary conditions and the Bondi gauge as above, we may solve for $\mu_2$.

In order to find the commutator, $\left[\xi_1, \xi_2 \right]$ of two generators we must repeat the process---acting first with $\xi_2$ and then with $\xi_1$, and find the corresponding values of $\mu_1$. We can then compute,
\eq{\delta \mu = \mu_2 - \mu_1\,,}
which gives the necessary piece that must be subtracted from the ordinary commutator to account for changes to the metric from the vector fields being themselves metric-dependent. 

It turns out that the only commutators for which this modification is important are those involving $T$. In \mbox{Appendix \ref{appendix:modified_bracket}} we illustrate this modified bracket in the most subtle case---showing that the commutator of two supertranslations, $\left[T, T \right]$, vanishes.

\vspace{-2pt}\subsection{The Conformal BMS Algebra}\label{subsec:conformal_bms_algebra}\vspace{-2pt}
In order to get a sense of the general structure of the group, it is useful to look at the elements involved in the commutation relations. The general results take the following overall form,
\eqs{\left[T , R \right] &\sim T\,, 
\\ \left[T, D \right] &\sim T\,, 
\\ \left[R, R \right] &\sim R\,,
\\ \left[C, D \right] &\sim C\,, 
\\ \left[D, E \right] &\sim E\,, 
\\ \left[E, R \right] &\sim C\,.\label{commutators_for_conformal_bms}}
We also have that 
\vspace{-5pt}\eq{\left[R, C\right] \sim E\,,\vspace{-5pt}}
except in the special case where the vector, $Y^A$ that generates the superrotations is a Killing vector, {\it i.e.}, $\psi = 0$, in which case,
\vspace{-5pt}\eq{\left[R, C\right] \sim C\,.\vspace{-5pt}}
All other commutators vanish:
\eqs{\left[T , T \right] &= 0\,, 
\\ \left[C, T \right] &= 0\,, 
\\ \left[C, C \right] &= 0\,,
\\ \left[R, D \right] &= 0\,, 
\\ \left[T, E \right] &= 0\,, 
\\ \left[C, E \right] &= 0\,, 
\\ \left[E, E \right] &= 0\,, 
\\ \left[D, D \right] &= 0\,.\label{vanishing_commutators_for_conformal_bms}}

One can now compare this algebra with that of the flat space conformal group. The first thing to notice is that the structure is entirely different. In particular, no commutator ever produces a dilatation on the right hand side. In the case of flat space, a special conformal transformation commuted with a translation gives a combination of dilatations and rotations. In this conformal BMS group, the commutation of both C and E with a supertranslation give zero. In addition, when a BMS special conformal transformation is commuted with a superrotation that is generated by a Killing vector, the result is consistent with the flat space version: we get another BMS special conformal transformation. However, when the superrotation is generated by a conformal Killing vector then the commutator gives a different result, a BMS dilatation. 

Both the flat space conformal group and the conformal BMS group have a subgroup involving the elements $T, R, D$, and these subgroups have the same structure---as seen in the first three lines of (\ref{commutators_for_conformal_bms}). The superrotations form their own subgroup, just like the rotations in the flat space group. 

Other subgroups of the conformal BMS group can be identified. There is one involving $T, D, E$, one with $E, R, C$, and one with $T, R$. There is another involving all elements except for the supertranslations, $R, C, D, E$. A dilatation with any other element also generates a subgroup.

With this group structure in mind, we can now look at the commutation relations in more detail. The supertranslations are generated by the function $f$, so we write \mbox{$T\!=\!T(f)$}. Similarly, the superrotations and special conformal transformations are generated by vector fields, so we write \mbox{$R\!=\!R(Y^A)$} and \mbox{$C\!=\!C(Z^A)$}. Then, more explicitly, the group algebra is given by,
\vspace{-5pt}\eq{\begin{array}{rl@{$\qquad$}rl}\fwboxR{95pt}{\left[T(f),D\right]}&\fwboxL{65pt}{=T(f')\,,}&\fwboxR{30pt}{f'}&\fwboxL{150pt}{= f\,,}\end{array}\label{td_commutator_detailed}\vspace{-26pt}}
\eq{\begin{array}{rl@{$\qquad$}rl}\fwboxR{95pt}{\left[T(f),R(Y^A)\right]}&\fwboxL{65pt}{=T(f')\,,}&\fwboxR{30pt}{f'}&\fwboxL{150pt}{= \frac{1}{2}f\,\psi - Y^A D_A f\,,}\end{array}\label{tr_commutator_detailed}\vspace{-20pt}}
\eq{\begin{array}{rl@{$\qquad$}rl}\fwboxR{95pt}{\left[D,C(Z^A)\right]}&\fwboxL{65pt}{= C((Z')^A)\,,}&\fwboxR{30pt}{(Z')^A}&\fwboxL{150pt}{= Z^A\,,}\end{array}\label{dc_commutator_detailed}\vspace{-20pt}}
\eq{\begin{array}{rl@{$\qquad$}rl}\fwboxR{95pt}{\left[R(Y^A), E  \right]}&\fwboxL{65pt}{=C((Z')^A)\,,}&\fwboxR{30pt}{(Z')^A}&\fwboxL{150pt}{= Y^A\,,}\end{array}\label{re_commutator_detailed}\vspace{-20pt}}
\eq{\begin{array}{rl@{$\qquad$}rl}\fwboxR{95pt}{\left[R(Y^A) , R((Y')^A)\right]}&\fwboxL{65pt}{=R((Y'')^A)\,,}&\fwboxR{30pt}{(Y'')^A}&\fwboxL{150pt}{= Y^B D_B (Y')^A - (Y')^B D_B Y^A\,.}\end{array}\vspace{-0pt}\label{rr_commutator_detailed}}
When $R$ is generated by a strict conformal Killing vector,
\eq{\left[R(Y^A),C(Z^A)\right]= \frac{1}{4} (\zeta \psi + D^A \zeta D_A \psi) E\,,}
whereas when $R$ is generated by a Killing vector,
\eq{\left[R(Y^A),C(Z^A)\right]= C((Z')^A),\qquad (Z')^A = Y^A \zeta\,.}
At first sight, when $R$ is generated by a strict CKV it does not look as though the commutator with $C$ gives simply $E$. However, closer inspection of the prefactor reveals that it is indeed a constant. This requires the following identities that hold for a 2$d$ strict CKV:
\eqs{Y^A &= - \frac{1}{2}D^A \psi\,,
\\ D_A D_B \psi &= - \gamma_{AB} \psi\,.}

Note that since the generators of $C$ must be strict conformal Killing vectors, equation (\ref{re_commutator_detailed}) shows that if the superrotation involved is generated by a Killing vector, then the commutator vanishes. While equation (\ref{rr_commutator_detailed}) gives a general expression for the commutation of two superrotations, it is worthwhile examining the result for the different cases in which the superrotations are generated by two KVs, two strict CKVs, or one of each. For either two KVs or two strict CKVs, the resulting superrotation generator, $(Y'')^A$ is a KV, but for one KV and one strict CKV, one gets a strict CKV.

We have checked all the Jacobi identities, and provide an illustrated example of how these commutation relations are computed according to the modified bracket in \mbox{Appendix \ref{appendix:modified_bracket}}.

\vspace{-8pt}\section{Conclusions and Discussion}\label{sec:conclusions}\vspace{-8pt}
The symmetries of spacetime at asymptotic infinity---especially in the case of asymptotically flat geometry---are of particular interest to the physics of scattering processes. In particular, this is where the $S$-matrix should be measured. The fact that there are more symmetries at infinity than mere Poincar\'{e} is extremely suggestive, and the connection between the holomorphically extended BMS group and the recently proposed infinite-dimensional symmetries of soft-particle scattering amplitudes \cite{Strominger:2013lka} related to soft-theorems \cite{Cachazo:2014fwa,Lysov:2014csa,Cheung:2016iub} may hint at a previously overlooked simplicity in the structure of four dimensional theories involving massless particles. 

Because many of the most intriguing results along these lines have been found in the context of the scattering of massless particles, the extension of the BMS group to include spacetimes with conformal symmetry is both natural and important. This is what we have done here. Continuing this generalization to the case of conformal theories with maximal supersymmetry is a natural road ahead---with exciting possibility of connecting the new symmetries proposed in \cite{Strominger:2013lka} with those known to exist in the case of maximally supersymmetric Yang-Mills theory in the planar limit. In a subsequent paper, a twistor representation of this group along with its supersymmetric extension will be discussed.

\vspace{\fill}
\vspace{-0pt}\section*{Acknowledgements}\vspace{-6pt}
This work was supported in part by the Danish National Research Foundation (DNRF91), a MOBILEX research grant from the Danish Council for Independent Research and a grant from the Villum Fonden (JLB), by the Avery-Tsui Foundation (SWH) and by STFC (SJH, MJP) and Trinity College research grants (MJP). We are also grateful to the support from the Cynthia and George Mitchell Foundation.

\appendix
\vspace{-6pt}\section{The Modified Bracket}\label{appendix:modified_bracket}\vspace{-8pt}
As explained in section 4.1, computing the algebra of the conformal BMS group required a delicate examination of the effect of each vector field on the spacetime, and how this would affect the action of a subsequent transformation. Here is a worked example for the commutator of two different supertranslations, $\left[T_1 , T_2 \right] \!=\! 0$.

Start by considering the action of a supertranslation, 
\eq{T_1 = g\,\partial_u + \frac{1}{2}D^2g\,\partial_r - \frac{1}{r}D^A g\,\partial_A\,.}
The ordinary commutator of this supertranslation, together with another supertranslation, $T_2$, generated by the function $f$, gives,
\eqs{\left[T_1, T_2 \right] &= \left[g\,\partial_u + \frac{1}{2}D^2g\,\partial_r - \frac{1}{r}D^A g\,\partial_A, f\,\partial_u + \frac{1}{2}D^2f\,\partial_r - \frac{1}{r}D^A f\,\partial_A\right], \\ &=\phantom{+} \frac{1}{2r}(D^Af D_A D^2g - D^Ag D_A D^2f)\,\partial_r \\&\phantom{=} + \frac{1}{2r^2}(D^2g D^Af - D^2f D^Ag + 2 D^Bg D_B D^A f - 2 D^B f D_B D^A g)\partial_A\,.}
This has the form,
\eq{\left[T_1, T_2 \right] = \frac{1}{r}A\,\partial_r + \frac{1}{r^2}B^A\partial_A\,,}
where $A$ and $B$ are functions of the two-sphere only. 

By considering dimensions, this implies that 
\eq{\mu_2^u = 0\,,}
and letting
\eq{\mu_2^r = \frac{1}{r}\hat{A}\,,}
and
\eq{\mu_2^A = \frac{1}{r^2}\hat{B}^A \,.}

Under the action of the first supertranslation the resulting infinitesimal changes to the metric are given by,
\eqs{\hat{h}_{uA} &= - \frac{1}{2} D_A(2 g + D^2 g)\,,
\\ \hat{h}_{AB} &= - r (2 D_A D_B g - \gamma_{AB} D^2g)\,,}
with all other components zero.

Then, under the action of the second supertranslation, $T_2$, on the metric there will be extra second order terms, $\hat{K}_{ab}$, given by,
\eqs{\hat{K}_{ab} =&\phantom{\,-\,}\mu_2^c\,\partial_c g_{ab} + T_2^c\,\partial_c \hat{h}_{ab} +\,\partial_a T_2^c \hat{h}_{bc} +\,\partial_a \mu_2^c g_{bc} +\,\partial_b \mu^c g_{ac} +\,\partial_b T_2^c \hat{h}_{ac} \\& - \frac{1}{2}g_{ab}\,\partial_c \mu_2^c - \frac{1}{2}\hat{h}_{ab}\,\partial_c T_2^c - \frac{1}{2}g_{ab} \Gamma^c_{\phantom{c}cd} \mu^d - \frac{1}{2}\hat{h}_{ab} \Gamma^c_{\phantom{c}cd} T_2^d - \frac{1}{2}g_{ab} \delta \Gamma^c_{\phantom{c}cd} T_2^d\,. }
The relevant Christoffel symbols and perturbations are given by,
\eqs{\Gamma^A_{\phantom{A}Ar} &= \frac{2}{r}\,,
\\\delta \Gamma^r_{\phantom{r}rA} &= \frac{1}{2r}D_A(D^2 + 2)g\,,
\\ \delta \Gamma^A_{\phantom{A}AB} &= - \frac{1}{2r} D_B(D^2+2)g\,.}
Thus, explicitly calculating the second order changes to the metric,
\eqs{\hat{K}_{rA}=0 &=g_{ru} D_A \mu^u + g_{AB}\,\partial_r \mu^B +\,\partial_r T_2^B \hat{h}_{AB}, \\ &=  - r^2 \hat{\gamma}_{AB} \left( \frac{2}{r^3}\hat{B}^A \right) - \frac{1}{r} D^B f ( 2 D_A D_B g - \gamma_{AB} D^2 g)\,,}
Therefore,
\eq{ \hat{B}_A = -\frac{1}{2}D^Bf (2 D_A D_B g - \gamma_{AB} D^2g)\,.}
\eq{\begin{split}
\hspace{-10pt} \hat{K}_{AB}=\mathcal{O}(r)&=\phantom{+}r^2(D_A \mu_B + D_B \mu_A - \frac{1}{2}\gamma_{AB}(\partial_u \mu^u +\,\partial_r \mu^r + D_C \mu^C - \frac{2}{r} \mu^r))\\&\phantom{=}+ D_A T_2^C \hat{h}_{BC} + D_B T_2^C \hat{h}_{AC} + D_A T_2^u \hat{h}_{uB} + D_B T_2^u \hat{h}_{uA} + T_2^r\,\partial_r \hat{h}_{AB} \hspace{-5pt}\\&\phantom{=} + T_2^C D_C \hat{h}_{AB} - \frac{1}{2} \hat{h}_{AB} D_C T_2^C  - \frac{1}{r}\hat{h}_{AB} T_2^r - \frac{1}{2}r^2 \gamma_{AB} \delta \Gamma^c_{\phantom{c}cd} T_2^d\,.
\end{split}}
Since
\eq{\hat{h}^A_A = 0\,,}
then,
\eqs{\hspace{-10pt}\hat{K}^A_A &= r^2 D_A \mu^A - r^2\,\partial_r \mu^r + 2r \mu^r + 2 D^A T_2^B \hat{h}_{AB} + 2 D^A T_2^u \hat{h}_{uA} - r^2 \delta \Gamma^c_{\phantom{c}cd} T_2^d\,,\hspace{-10pt} \\ &= 2r \mu^r -r^2\,\partial_r \mu^r + r^2 D_A \mu^A +2D^AD^Bf(2D_AD_Bg-\gamma_{AB}D^2g)\\
&\phantom{=}-D^Af D_A(D^2+2)g,\\
&=3\hat{A}-\frac{1}{2}D^AD^Bf(2D_AD_Bg-\gamma_{AB}D^2g)-\frac{1}{2}D^Bf(2D^2D_Bg-D_BD^2g)\\
&\phantom{=}+2D^AD^Bf(2D_AD_Bg-\gamma_{AB}D^2g)-D^AfD_A(D^2+2)g,\\
& =  3\hat{A} + 3 D^A D^B f D_A D_B g - \frac{3}{2} D^2 f\, D^2g  - \frac{3}{2} D^B f\,D_B D^2g -3 D^A f D_A g\,.}

Since
\eq{\partial_r \left(\frac{\det(g_{AB})}{r^2} \right) = 0\,,}
we have,
\eq{\hat{A} =  \frac{1}{2} D^B f D_B D^2g - D_A D_B f D^A D^B g + \frac{1}{2}D^2 f D^2g + D^A f D_A g\,.}
Thus,
\eqs{\mu_2^u &= 0\,,
\\ \mu_2^r &= \frac{1}{r} \left(\frac{1}{2} D^B f D_B D^2g - D_A D_B f D^A D^B g + \frac{1}{2}D^2 f D^2g +D^A f D_A g \right)\,,
\\ \mu_2^A &= - \frac{1}{2 r^2} D^Bf (2 D_A D_B g - \gamma_{AB} D^2g)\,.}

When we perform the same set of calculations using first the action of $T_2$, followed by $T_1$, we get the same results for $\mu_1^a$, with $f \!\leftrightarrow\!g$.

Therefore, we can calculate,
\eq{\delta \mu^a = \mu_2^a - \mu_1^a\,,}
to find,
\eqs{\delta \mu^u  &= 0\,,
\\ \delta \mu^r &= \frac{1}{2r}(D^B f D_B D^2 g - D^B g D_B D^2 f)\,,
\\ \delta \mu^A &= \frac{1}{2r^2} (D_Bg (2 D^A D^B f - \gamma^{AB} D^2f) - D_Bf (2 D^A D^B g - \gamma^{AB} D^2g)) \\ &= \frac{1}{2r^2}(D^2g D^Af - D^2f D^Ag + 2 D^Bg D_B D^A f - 2 D^B f D_B D^A g)\,.}
These terms exactly cancel those arising from the ordinary commutator, and so upon subtracting these off, we find that,
\eq{\left[T_1, T_2 \right] = 0\,.}


\newpage
\providecommand{\href}[2]{#2}\begingroup\raggedright\endgroup

\end{document}